\documentclass[12pt]{article}
\usepackage{times}
\usepackage{geometry}
\usepackage[utf8]{inputenc}
\usepackage{enumitem,amssymb}
\usepackage{ragged2e}
\usepackage{graphicx}
\newlist{thematic}{itemize}{8}
\setlist[thematic]{label=$\square$}
\usepackage{pifont}

\usepackage[numbers]{natbib}
\usepackage{float}
\usepackage{caption}
\usepackage{placeins}

\begin{document}
\raggedright
\huge

Extending Hubble into the 2030s to Resolve the Physics of LyC Escape \linebreak
\normalsize

  
\textbf{Principal Author:} Cody Carr (University of Michigan) \texttt{codycarr@umich.edu} 

\justifying{\noindent
\textbf{Co-authors:} S. R. McCandliss (Johns Hopkins University), M. A. Berg (Texas Christian University), Renyue Cen (Zhejiang University), Kevin France (University of Colorado Boulder), Matthew J. Hayes (Stockholm University), Alaina Henry (Space Telescope Science Institute), M. S. Oey (University of Michigan), Alberto Saldana-Lopez (Stockholm University)\ \ \ \ \ \ \ \ \ \ \ \ }
  \linebreak

\textbf{Abstract:}
\justifying
Current observations with the James Webb Space Telescope (JWST) suggest that star-forming galaxies produce enough ionizing (LyC; $\lambda < 91.2$ nm) photons to drive cosmic reionization, but the efficiency with which these photons escape their host galaxies remains uncertain. Absorption by the neutral intergalactic medium progressively suppresses direct LyC detections above redshift $z\sim3$, forcing astronomers to rely on indirect diagnostics of LyC escape calibrated at low redshift. Low-resolution ultraviolet observations of high-redshift analogs obtained with the Cosmic Origins Spectrograph onboard the Hubble Space Telescope (HST) have been critical for developing these diagnostics. These studies suggest that stellar feedback plays a central role in regulating LyC escape, although the role of galactic winds and the underlying physical mechanisms remain poorly constrained. High-resolution spectroscopy blueward of 1600~\AA\ (rest-frame) is required to resolve the kinematic structure of the winds and reveal the physics governing LyC escape. Such observations are currently only possible with HST and represent a major science driver for the future Habitable Worlds Observatory (HWO). Extending the lifetime of HST and prioritizing ultraviolet observations are essential for interpreting current JWST studies of the early Universe and important preparatory science for HWO.   

\section{Introduction}

The Epoch of Reionization (EoR) represents the last major phase transition of the Universe, during which ionizing or Lyman continuum (LyC; $\lambda < 91.2$ nm) photons emitted by the first galaxies ionized the majority of neutral hydrogen filling the intergalactic medium (IGM). Many fundamental questions regarding the EoR remain unanswered. In particular, the dominant sources of reionization remain uncertain, limiting our understanding of how the earliest cosmic structures reionized the Universe.  

Current observations with the James Webb Space Telescope (JWST) suggest that star-forming galaxies produce enough LyC photons to drive cosmic reionization, although the efficiency with which these photons escape their host galaxies—and the relative contribution of other sources such as AGN—remain uncertain \cite{Finkelstein2023,Sanders2023,Llerena2024,Simmonds2024,Lin2024,Munoz2024,Pahl2025}. Absorption by the neutral IGM progressively suppresses direct LyC detections above redshift $z\sim3$ \cite{Inoue2014}, forcing astronomers to rely on indirect diagnostics of LyC escape calibrated at low redshift.  For example, the largest of these studies has been the Low-z LyC Survey+ (LzLCS+), confirming 50 LyC emitters among 89 galaxies \cite{Flury2022_data,Flury2022_diagnostics}.  

LzLCS+ examined several proposed indicators of LyC escape, including star formation rate surface density ($\Sigma_{\rm SFR}$; \cite{Heckman2001,Heckman2011,Clarke2002,Zastrow2011,Sharma2017,Naidu2020}), high [O III]/[O II] ratios (O32; \cite{Jaskot2013,Nakajima2014}), and the FUV continuum slope ($\beta$; \cite{Zackrisson2013,Zackrisson2017,Chisholm2022}), but found that no single parameter robustly predicts $f_{\rm esc}^{\rm LyC}$. This result implies a complex interplay between the timing of LyC production, stellar feedback, and the H I and dust content of the interstellar medium (ISM) and surrounding circumgalactic medium (CGM) regulating LyC escape \cite{Flury2022_data}. Multivariate approaches show greater promise, but their predictions differ by several orders of magnitude at high redshift, both among themselves and relative to simulation-based models \cite{Mascia2023,Choustikov2024,Mascia2024,Jaskot2024b,Jaskot2025}. To determine how efficiently star-forming galaxies contributed to the reionization of the Universe, we must first establish the detailed physics governing LyC escape at low redshift. Only then can we understand how these processes evolve with redshift. The challenge of this problem lies in its multiscale nature: LyC photons are produced in stellar clusters (10–100 pc) but must ultimately escape through the CGM (10–100 kpc).

Stellar feedback—including radiation, supernovae (SNe), and stellar winds—has long been recognized as a key requirement for clearing pathways for LyC escape in star-forming galaxies \cite{Heckman2001}. However, the dominant feedback mechanism, the timing of LyC escape, and the role of galactic winds in clearing H I and dust from the ISM and CGM remain uncertain. Cosmological simulations have long favored SN-driven outflows as the primary mechanism for opening LyC escape channels through the ISM, but 
typically only after the most massive LyC-producing stars have already died ($>10$ Myr; \cite{Kimm2014,Ma2020,Rosdahl2022,Choustikov2024}). Alternatively, theoretical arguments and turbulent molecular cloud simulations suggest that radiation can evacuate or ionize low-column-density channels at much earlier times ($<3$ Myr), both with fast ($\sim$100s km s$^{-1}$; \cite{Menon2025,Ferrara2025}) and without fast ($\sim$10s km s$^{-1}$; \cite{Kakiichi2021}) winds. \citet{Jecmen2023} point out that mechanical feedback is not expected to generate strong galactic superwinds until $\sim 10$ Myr in metal-poor galaxies since SNe are not expected to dominate until $\sim20\ M_\odot$ stars expire, suggesting that radiation-driven feedback must facilitate LyC escape at the earliest times.  The timing is especially critical for constraining a galaxy’s contribution to reionization, as the intrinsic production of LyC photons peaks early while the most massive stars are still alive. Consequently, the total escaping LyC emissivity depends sensitively on the overlap between intrinsic LyC production and the escape fraction ($f_{\rm esc}^{\rm LyC}$).

Observations support the existence of both mechanical and radiation-dominated feedback. The strongest LyC emitters tend to exhibit high O32 ratios \cite{Flury2022_data}, lack evidence for significant SN activity in radio emission \cite{Bait2024}, and host young stellar populations \cite{Carr2025LyC,Flury2025}, favoring a radiation-driven scenario. 
Green pea galaxies with the most extreme O32 ratios also lack kinematic evidence of mechanical feedback and SN activity \cite{Carr2025Sup, Jaskot2017}, while slightly older objects support their presence \cite{Carr2025Sup, Komarova2025}.  Low-resolution spectral stacks suggest that strong LyC emitters host both young and old stellar populations, pointing toward a confluence of radiative and SN feedback operating together in a two-stage burst scenario \cite{Flury2025}. Emission-line widths from ionized gas (e.g., [O III]) correlate with $f_{\rm esc}^{\rm LyC}$, suggesting the presence of fast winds that could be driven by either SNe or radiation \cite{Amorin2024,Komarova2025}. In contrast, low-ionization metal tracers (e.g., Si II) generally do not show similar trends \cite{Chisholm2017a,Jaskot2017,Carr2025LyC}, possibly indicating a phase dependence.

These studies are typically limited by low spectral resolution and/or rely on the use of indirect metal tracers of H I, often leading to conflicting physical interpretations of the same galaxies (cf. \cite{Carr2025LyC,Li2025}). Constraining the properties of galactic winds requires high-resolution, high signal-to-noise (S/N) spectroscopy capable of resolving wind kinematics \cite{Carr2023,Carr2025MOR} and distinguishing stellar absorption from galactic \cite{Carr2025Sup}. This critical region of parameter space remains largely unexplored and is currently accessible only with the Hubble Space Telescope (HST). Extending HST operations into the 2030s will therefore be essential for establishing the detailed physics governing LyC escape ahead of the future ultraviolet Habitable Worlds Observatory (HWO).

\section{Science Objectives}

\begin{itemize}[noitemsep]
    \item How does the LyC escape fraction evolve alongside the baryon cycle?
    \item What is the average escaping LyC emissivity produced by a galaxy over its baryon cycle?
    \item What is the primary driver of LyC escape: radiation or SNe?  When does each dominate?
    \item How does radiative feedback enable LyC escape: ionization or pressure-driven winds?
    \item How do the properties of galactic winds, inferred from absorption-line profiles, relate to the physics of LyC escape and commonly used escape diagnostics (e.g., O32, $\beta$, $\Sigma_{\rm SFR}$)?  How do these relationships evolve with redshift? 
\end{itemize}

\section{Pilot Study HST-GO-17433}

 \begin{figure}
	\centering 
\includegraphics[width=\textwidth]{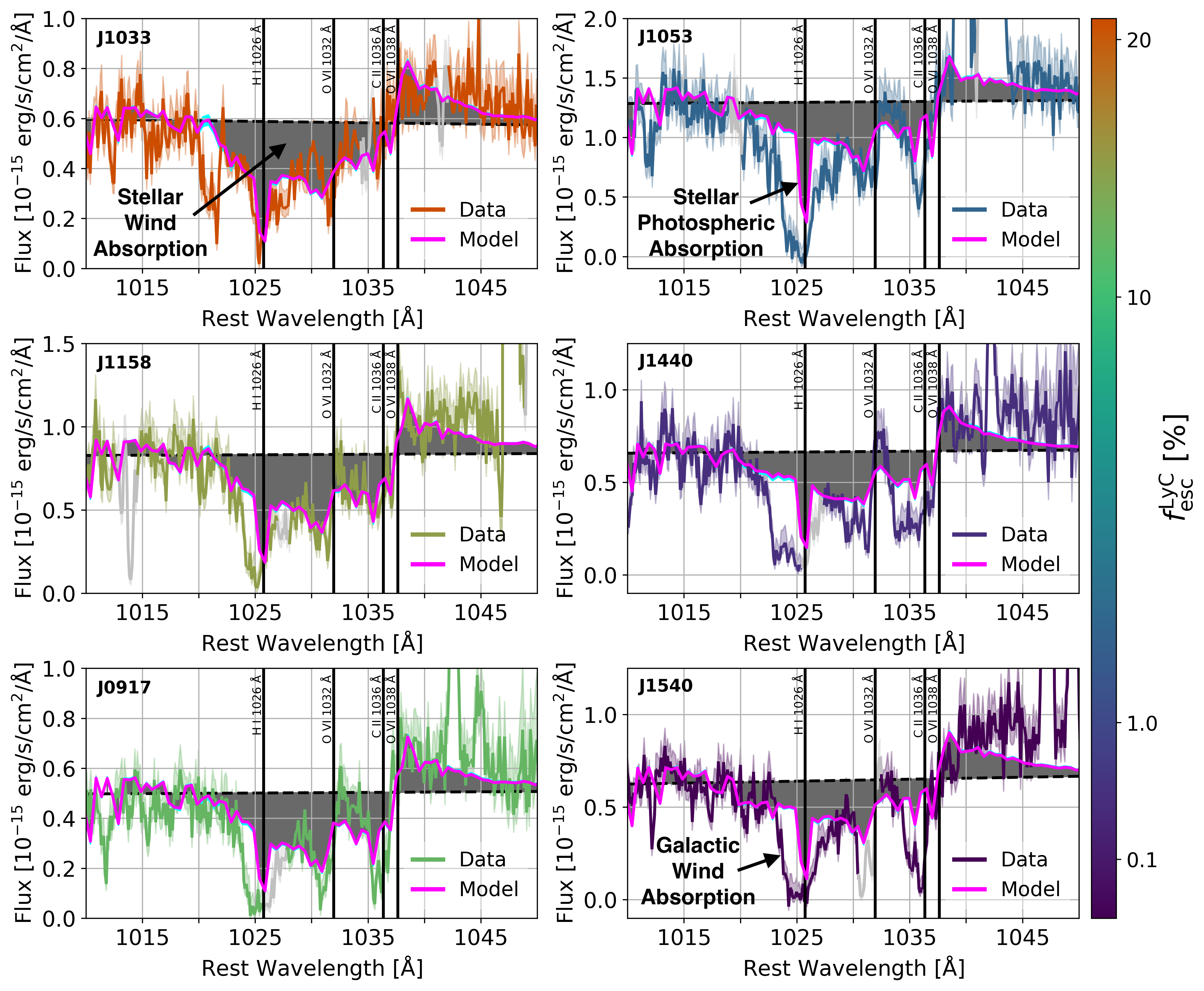}
    \vspace{-0.75cm}
    \caption{\justifying \textbf{Stellar wind diagnostics from HST-GO-17443 \cite{Carr2025Sup}.} H\,I $\lambda1026$, C\,II $\lambda1036$, and O\,VI $\lambda\lambda1032,1038$ profiles for galaxies ordered by decreasing $f_{\rm esc}^{\rm LyC}$ from top to bottom, left to right, according to the color bar. Best-fitting stellar SED models, which reproduce the stellar photospheric and stellar-wind absorption features, are shown in magenta with absorption regions shaded in grey. Additional absorption, unaccounted for by the SED model, is attributed to galactic winds.}
	\label{fig:stellar_winds}
\end{figure}

 \begin{figure}
	\centering 
	\includegraphics[width=\textwidth]{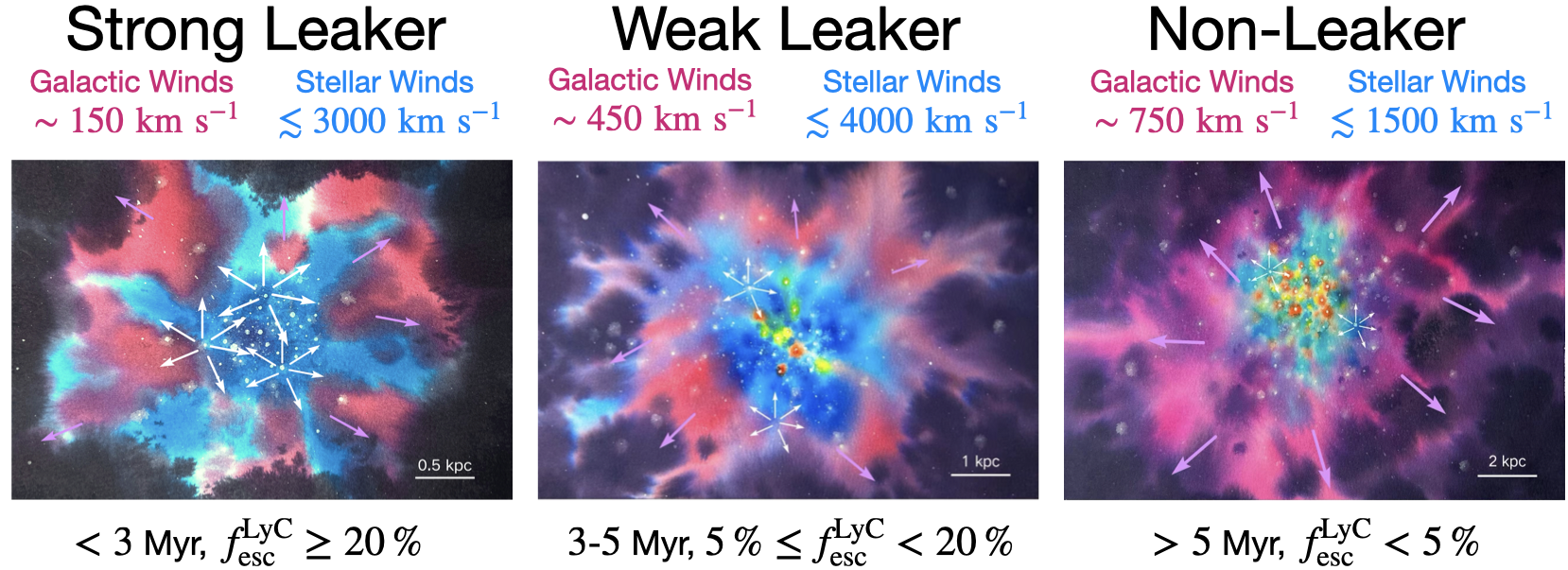}
    \vspace{-0.75cm}
    \caption{\justifying \textbf{LyC escape sequence for HST-GO-17443 \cite{Carr2025Sup}.} Early radiative feedback ($\sim$3 Myr) opens low-column-density channels and drives diffuse winds, yielding high $f_{\rm esc}^{\rm LyC}$. As SNe emerge (3–5 Myr), continued feedback expels dense gas that ultimately suppresses LyC escape ($\gtrsim$5 Myr). Additional COS/HST observations spanning a broader range of galaxy properties are required to test the statistical significance of this evolutionary sequence.}
	\label{fig:LyC_Escape_Sequence}
\end{figure}

Program HST-GO-17433 obtained the first sample of high resolution ($\sim 75\ \rm km\  s^{-1}$), high signal-to-noise (S/N $=10$ at 1000\AA) data of confirmed LyC emitters to capture the structure of their winds. 
The pilot program obtained COS G130M spectra of six LzLCS+ galaxies, directly targeting H I in absorption through the Lyman series \cite{Carr2023HST}.  Figure~\ref{fig:stellar_winds} shows the H I $\lambda1026$, C II $\lambda1036$, and O VI $\lambda\lambda1032,1038$ absorption lines for each galaxy ordered from top to bottom, left to right with decreasing $f_{\rm esc}^{\rm LyC}$.  Each spectrum was fit using the FiCUS spectral energy distribution (SED) code\cite{Saldana-Lopez2022}, which relies on starburst99 models\cite{Leitherer1999,Leitherer2014} to capture stellar wind absorption and photospheric absorption.  The high resolution of the study allows for a separation of galactic and stellar absorption components, where galactic absorption is not captured by the SED.  The decrease in $f_{\rm esc}^{\rm LyC}$ correlates almost perfectly with the mean light weighted age of the stellar populations, suggesting that the decline in $f_{\rm esc}^{\rm LyC}$ is tracing the launch of a galactic wind \cite{Carr2025Sup}.       


Together, these results suggest that LyC escape peaks early and is later suppressed by the launch of a multiphase galactic wind. The average stellar age of J1033—the strongest LyC emitter—is $2.71\pm0.75$ Myr, suggesting that the wind in this galaxy was launched at even earlier times \cite{Carr2025Sup}. This timing favors a scenario in which radiation and stellar winds, rather than SNe, drive the outflow and create optimal conditions for LyC escape in J1033.  These results are summarized in Figure~\ref{fig:LyC_Escape_Sequence}.      

\section{Required Observations with HST}

\begin{figure}
	\centering 
	\includegraphics[width=\textwidth]{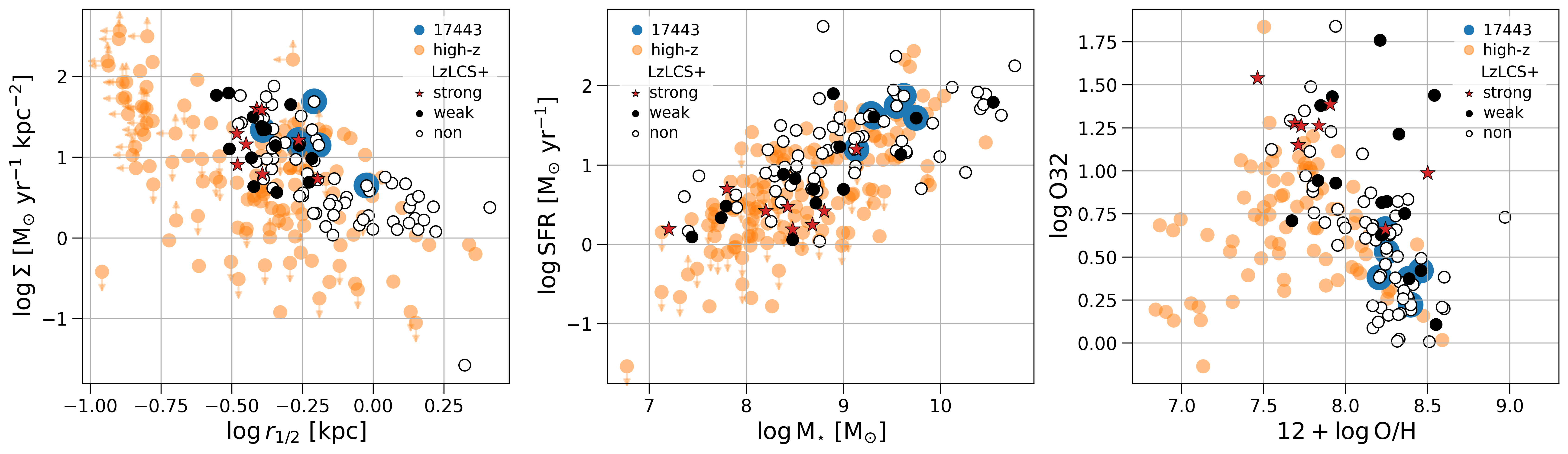}
    \vspace{-0.75cm}
    \caption{\justifying \textbf{Comparison of HST-GO-17443 (blue) samples with LzLCS+ and $\mathbf{z>5}$ galaxies from CEERS and GLASS \cite{Calabro2024} (orange).} LzLCS+ galaxies are classified as strong ($f_{\rm esc}^{\rm LyC} > 20\%$, red stars), weak ($5 < f_{\rm esc}^{\rm LyC} < 20\%$, black circles), and non-leakers ($f_{\rm esc}^{\rm LyC} < 5\%$, white circles). }
	\label{fig:low_vs_high_redshift}
\end{figure}

We compare the properties of the HST-GO-17443 galaxies to the broader LzLCS+ sample and to $z>5$ galaxies observed with the James Webb Space Telescope (JWST) in CEERS and GLASS \cite{Calabro2024} in Figure~\ref{fig:low_vs_high_redshift}. The HST-GO-17443 galaxies were selected to be nearly identical in stellar mass ($9.13 \leq \log M_{\star}\ [{\rm M}_\odot] \leq 9.75$), SFR ($1.3 \leq \log{\rm SFR}\ [{\rm M}_{\odot}\ {\rm yr}^{-1}] \leq 1.6$), and metallicity ($8.2 \leq 12+\log{\rm O/H} \leq 8.5$). Importantly, the galaxies are also compact ($R_{\rm UV}<1.0$ kpc) and exhibit high $\Sigma_{\rm SFR}$ ($\log \Sigma \ [\rm M_{\odot}\ yr^{-1}\ kpc^2]>0.5$), properties associated with efficient stellar feedback and enhanced LyC escape \cite{Cen2020}. As a result, the current sample probes only a limited region of the LzLCS+ parameter space, leaving open important questions regarding how the LyC escape sequence shown in Figure~\ref{fig:LyC_Escape_Sequence} evolves toward lower metallicities and other galaxy properties. A larger and more diverse sample is therefore required to establish a more complete physical picture in the local Universe. Naturally, galaxies at high redshift are expected to span an even broader range of parameter space.

HST-GO-17443 required a sensitivity of $S/N_{1000}^{5,\rm res}=10$ at 1000~\AA, together with a spectral resolution of $\sim75\ \rm km\ s^{-1}$ ($\approx0.25$\AA; 5 resels). This setup was selected to optimize the detectability of saturation in the cores of the H I absorption lines while resolving outflow kinematics on scales better than $\sim100\ \rm km\ s^{-1}$, substantially above the limiting G130M resolution element of $\sim15\ \rm km\ s^{-1}$ ($\approx0.05$\AA). The resulting LyC detections achieved S/N $\approx5$ when rebinned to a resolution of 2~\AA\ (40 resels). These observations are observationally expensive, in some cases requiring more than 10 HST orbits per target. Extending this analysis across the full LzLCS+ parameter space ($\sim$ 20 galaxies) would require a dedicated deep survey totaling around 200 orbits. 

Extending HST operations into the 2030s to obtain deep FUV spectroscopy will be critical for establishing the detailed physics governing LyC escape and preparing for the next generation of ultraviolet observatories such as HWO. In particular, HST/COS provides the only currently available platform capable of directly probing the multiphase wind structure of confirmed LyC emitters at the spectral resolution and sensitivity needed to disentangle stellar feedback, neutral gas structure, and galactic outflows. While upcoming ultraviolet missions such as SPRITE and UVEX will provide transformative wide-field ultraviolet imaging and spectroscopic capabilities, neither mission is optimized for the deep, high-resolution far-ultraviolet absorption-line spectroscopy needed to characterize the multiphase wind kinematics of faint LyC-emitting galaxies. Specifically, the limited aperture of SPRITE and the moderate spectral resolution of UVEX are insufficient to achieve the sensitivity and velocity resolution required to separate stellar absorption from outflowing gas and directly constrain the neutral ISM geometry governing LyC escape.  These observations will not only define the physical framework needed to interpret reionization-era galaxies observed with JWST, but will also provide essential guidance for HWO instrument development and mission design providing target wavelengths, sensitivities, and resolutions. Indeed, high-resolution LyC surveys have emerged as an HWO "Driving Science Case" (HWO CSIT; AAS 2026 Town Hall; private communication), motivating the development of ultraviolet multi-object and integral field spectroscopy capabilities for HWO \cite{Citro2025HWO,McCandliss2025HWO,Xu2025HWO,Carr2025JATIS,Carr2026}.

\pagebreak

\bibliographystyle{aasjournal_nodoi}
\bibliography{references}

\end{document}